
\magnification 1200
\hsize=6.0 true in
\hoffset=0.1 true in
\baselineskip 15pt


\def\r2{\sqrt 2}
\def\T3f{T_{3f}}
\def\sw2{\sin^2\theta_W}
\def\stw{\sin 2\theta_W}

\def\ctw{\cot\theta_W}
\def\ct2w{\cot 2\theta_W}
\def\v#1{v_#1}

\def\tb{\tan\beta}
\def\s2b{\sin 2\beta}
\def\c2b{\cos 2\beta}
\def\s2b2{\sin^22\beta}
\def\csthR{\cos\theta_R}
\def\snthR{\sin\theta_R}
\def\csthL{\cos\theta_L}
\def\snthL{\sin\theta_L}

\def\w#1{\omega_#1}

\def\wi{\omega_i}

\def\mgr{m_{3/2}}
\def\mw{m_{\omega}}

\def\m#1{\tilde {m_#1}}
\def\mH{m_H}

\def\mw#1{m_{\omega #1}}

\def\GL12{G_L^{12}}
\def\GR12{G_R^{12}}

\font\bigit=cmmi10 scaled \magstep3

\font\bigbf=cmbx10 scaled \magstep3

\line{\hfill IFM 11/93}
\line{\hfill TKU-HEP 93/06}

\vskip 3.0 true cm
\centerline{\bigbf
{\bigit T}--odd Asymmetry}
\bigskip\centerline{\bigbf
in Chargino Pair Production Processes
\footnote{$^*$}{\rm to appear in the proceedings
of the Workshop on $e^+e^-$ Collisions at 500 GeV.}
}
\vskip 1.0 true cm
\centerline{Yoshiki Kizukuri$^1$ and Noriyuki Oshimo$^2$}
\vskip 1.0 true cm
\item{$^1$} {\it Department of Physics, Tokai University}
\item{ } {\it 1117 Kita-Kaname, Hiratsuka 259-12, Japan}

\item{$^2$} {\it Grupo Te\'orico de Altas Energias, CFMC}
\item{ }
{\it Av. Prof. Gama Pinto, 2, 1699 Lisboa Codex, Portugal}

\vskip 2.0 true cm\vfill
\centerline{\bigbf Abstract}\medskip

$T$--violating asymmetry in chargino pair production processes
is studied in the Minimal Supersymmetric Standard Model.
The asymmetry emerges
at the tree level in the production of two different charginos,
and could be
as large as of order $10^{-2}$ in unpolarized electron beam experiments
and $10^{-1}$ in polarized electron beam experiments.
In the pair production
of the same charginos, the asymmetry emerges through the electric and the weak
"electric" dipole moments of the charginos at the loop level.  Its magnitude
is of order $10^{-4}$.  The consistency with the electric dipole
moments of the neutron and the electron is also discussed.

\smallskip
\eject

\noindent{\bigbf 1. Introduction}\medskip

Supersymmetry reduces the number of the parameters contained in a model.
However, once it is violated, the number of the parameters proliferates.
In an extension of the standard model with the exact supersymmetry,
we have only the Kobayashi-Maskawa phase as a source of $CP$--violation.
We know that the breakdown of the supersymmetry must occur at least
at the electro-weak energy scale to prevent the supersymmetric partners
of the known particles from appearing under the LEP and Tevatron energy scale.
In realistic extensions
of the standard model with the broken supersymmetry,
we have new complex parameters as sources of $CP$--violation in addition to
the Kobayashi-Maskawa phase.

It is well-known that these new complex
parameters give large contributions
to the electric dipole moments ( EDMs ) of the neutron [1-3] and
the electron [4], which would be taken as a troublesome problem of
the supersymmetric standard model. From the present experimental
upper bounds of the EDMs of the neutron [5] and the electron [6],
the phases of the complex parameters should be less than order
of $10^{-2}$ or the masses of the super-partners,
especially squarks and sleptons, should
be larger than order of 1 TeV [7].
We take the latter position in this report,
i.e. the phases of the complex parameters of the model are of order 1, but
super-partners are heavy enough to clear the EDM constraints.

These new $CP$--violating phases cause various $CP$-- and $T$--violating
phenomena [8-12].  Amongst them
$T$--odd phenomena may be possible
to be observed in high energy collider experiments,
especially in $e^+e^-$ collisions at 500 GeV.
In the previous report [11] we considered,
as an effect of $CP$--violation,
the $T$--odd asymmetry
in the neutralino production processes
by $e^+e^-$ annihilation.  We are now to discuss the chargino case.

\vskip 1.0 true cm
\noindent{\bigbf 2.  {\bigit CP}--violating Interactions}\medskip

The Lagrangian of the minimal supersymmetric extension of
the standard model (MSSM) [13] is given by
$$
    L = L_{kin} + L_{gauge} + L_F + L_S;
\eqno(1)
$$
$$\eqalign{
        L_F =& [ (E^c Y_E L)H_1 + (D^c Y_D Q)H_1
               + (U^c Y_UQ)H_2 + m_HH_1H_2 ]_F
               + {\rm h.c.}, \cr
      - L_S =& (\tilde E^c\eta_E\tilde L)\tilde H_1
                + (\tilde D^c\eta_D\tilde Q)\tilde H_1
                + (\tilde U^c\eta_U\tilde Q)\tilde H_2
                +  \tilde M^2_H\tilde H_1\tilde H_2 \cr
  &+ {1\over 2}\sum_{i=1}^3{\tilde m}_i\bar\lambda_{iR}
                                           \lambda_{iL}
   + {1\over 2}\sum_{a,b}\tilde M_{ab}^2\phi_a^*\phi_b
             + {\rm h.c.}.}
$$
If one assumes $N$=1 supergravity and
the grand unification of the fundamental interactions,
some of these parameters in (1) could be related to each other
at the unification scale,

$$\eqalign{
     \tilde M_{ab}^2 =& |\mgr|^2 \delta_{ab}, \cr
         \eta_f =& A\mgr Y_f\ \ \ \ \ \ \  ( f = E, D, U ), \cr
         \tilde M_H^2 =& B\mgr m_H, \cr
         {\tilde m}_i =& M_\lambda
               \ \ \ \ \ \ \ \ \ \ \ \ \  ( i = 1, 2, 3 ). \cr}
\eqno(2)
$$
Besides the Yukawa coupling constants $Y_f$
and the vacuum
expectation values of the two neutral Higgs scalar fields,
the complex parameters involved in the model are
$A\mgr$, $B\mgr$, $m_H$, and $M_\lambda$.

These complex parameters can in general become sources of $CP$--violation.
The $CP$--violating interactions appear through rewriting the interactions
by the mass eigenstates of the particles
obtained by the diagonalization of the mass matrices.
The mass matrix of our interest is the chargino one,
$$
( \bar{\lambda^-_R}, \bar{\psi_2^c} ) M^- \left(\matrix{\lambda^-_L \cr
                                                          \psi_1}\right)
                         + {\rm h.c.};
\eqno(3)
$$
$$
    M^- = \left(\matrix{\m2 & -g\v1^*/\r2 \cr
                -g\v2^*/\r2 & \mH}        \right),
$$
where $\v1$ and $\v2$ are the vacuum expectation
values of $\tilde H_1$ and $\tilde H_2$,
respectively.  Extracting the phases of $\m2$, $v_1$, and $v_2$ ,
we can write
$$
M^- = P_R M_C P_L^{\dagger};
\eqno(4)
$$
$$\eqalign{
      M_C &= \left(\matrix{|\m2|  & g|v_1|/\sqrt 2\cr
                   g|v_2|/\sqrt 2 & |\mH|e^{i\theta}\cr}\right),\cr
      P_R &= \left(\matrix{e^{i\theta_g} & 0 \cr
                    0 & -e^{-i\theta_2}}\right), \quad
      P_L = \left(\matrix{1      &  0 \cr
                    0 & -e^{i(\theta_1+\theta_g)}}\right), }
$$
where $\theta = \theta_H + \theta_g + \theta_1 + \theta_2$,
$\theta_H = \arg(\mH)$, $\theta_g = \arg(\m2)$,
$\theta_1 = \arg(v_1)$, and
$\theta_2 = \arg(v_2)$.  The complex $2\times 2$ matrix $M_C$
can be readily diagonalized by unitary matrices $U_{R, L}$ as
$$
    U_R^{\dagger}M_C U_L = \left( \matrix{ \mw1 & 0 \cr
                                             0    & \mw2}\right);
\eqno(5)
$$
$$\eqalign{
    &U_R =\left(\matrix{\csthR  & -e^{-i\beta_R}\snthR \cr
                    e^{i\beta_R}\snthR & \csthR       }\right)
          \left(\matrix{e^{i\gamma_1} & 0 \cr
                        0 & e^{i\gamma_2}}\right),               \cr
    &U_L =\left(\matrix{\csthL  & -e^{-i\beta_L}\snthL \cr
                    e^{i\beta_L}\snthL & \csthL       }\right),         \cr
}
$$
where the angles are calculated to be
$$ \eqalign{
\tan~2\theta_R &= \sqrt 2 g{\sqrt{|v_2\m2|^2+|v_1\mH|^2
                      +2|v_1v_2\m2\mH|\cos\theta}
               \over |\m2|^2 - |\mH|^2+ g^2(|v_1|^2 - |v_2|^2)/2},\cr
\tan~2\theta_L &= \sqrt 2 g{\sqrt{|v_1\m2|^2+|v_2\mH|^2
                      +2|v_1v_2\m2\mH|\cos\theta}
               \over |\m2|^2 - |\mH|^2+ g^2(|v_2|^2 - |v_1|^2)/2},\cr
\tan\beta_R  &= {\sin\theta \over\displaystyle \cos\theta +
                 {\strut|v_2\m2| \over\displaystyle |v_1\mH|}},
\quad
\tan\beta_L  = -{\sin\theta \over\displaystyle \cos\theta +
                 {\strut|v_1\m2| \over\displaystyle |v_2\mH|}},    \cr
\tan\gamma_1 &= -{\sin\theta\over\displaystyle \cos\theta +
             {\strut|\m2|\over\displaystyle|\mH|}
       {\strut2(\mw1^2 - |\mH|^2)\over\displaystyle g^2|v_1v_2|}}, \cr
\tan\gamma_2 &= {\sin\theta\over\displaystyle \cos\theta +
           {\strut|\m2|\over\displaystyle|\mH|}
         {\strut g^2|v_1v_2|\over\displaystyle 2(\mw2^2 - |\m2|^2)}}.\cr
}\eqno(6)
$$
We note that only one combination of the four complex phases, $\theta$,
acts as a physical $CP$--violation source.
Finally the chargino mass eigenstates $\wi$
and the unitary matrices $C_{R, L}$
diagonalizing (3) are obtained from $U_{R, L}$ and $P_{R, L}$ as
$$\eqalign{
\left(\matrix{{\w1}_L \cr
            {\w2}_L} \right) &= C_L^{\dagger}\left(\matrix{\lambda^-_L \cr
                                                    \psi_1}\right), \quad
\left(\matrix{{\w1}_R \cr
            {\w2}_R} \right) = C_R^{\dagger}\left(\matrix{\lambda^-_R \cr
                                                   \psi_2^c}\right), \cr
C_{R, (L)} &= P_{R, (L)}U_{R, (L)}.\cr
}\eqno(7)
$$

The neutral current of the charged gaugino $\lambda^-$ and
the charged Higgsinos $\psi_{1, 2}$,
$$
J^Z_{\mu} = - e\ctw\bar{\lambda^-}\gamma_{\mu}\lambda^-
           - e\ct2w\bar{\psi_1}\gamma_{\mu}\psi_1
           - e\ct2w\bar{\psi_2^c}\gamma_{\mu}\psi_2^c,
\eqno(8)
$$
is now rewritten by the mass eigenstates $\wi$ as
$$\eqalign{
J^Z_{\mu} = &e\bar{\w1}\gamma_{\mu}[ G_L^{11}P_- + G_R^{11}P_+ ]\w1
          + e\bar{\w2}\gamma_{\mu}[ G_L^{22}P_- + G_R^{22}P_+ ]\w2 \cr
          + &e\bar{\w1}\gamma_{\mu}[ G_L^{12}P_- + G_R^{12}P_+ ]\w2
          + e\bar{\w2}\gamma_{\mu}[ G_L^{21}P_- + G_R^{21}P_+ ]\w1, \cr
}\eqno(9)
$$
where $P_{\pm} = (1 \pm \gamma_5)/2$, and
$$\eqalign{
G_{L, R}^{11} &= - \ctw |C_{L, R}^{11}|^2 - \ct2w |C_{L, R}^{21}|^2, \cr
G_{L, R}^{22} &= - \ctw |C_{L, R}^{12}|^2 - \ct2w |C_{L, R}^{22}|^2, \cr
G_{L, R}^{12} &= - \ctw C_{L, R}^{11*}C_{L, R}^{12}
               - \ct2w C_{L, R}^{21*}C_{L, R}^{22}, \cr
G_{L, R}^{21} &= - \ctw C_{L, R}^{12*}C_{L, R}^{11}
               - \ct2w C_{L, R}^{22*}C_{L, R}^{21}. \cr
}\eqno(10)
$$
The coupling constants $G_{L, R}^{21}$ of
the off-diagonal current can be explicitly
given as
$$\eqalign{
G_L^{21} &= {\strut 1\over \displaystyle \stw} e^{i\beta_L} \csthL\snthL, \cr
G_R^{21} &= {\strut 1\over \displaystyle \stw}
             e^{i( \beta_R + \gamma_1 - \gamma_2 )} \csthR\snthR, \cr
}\eqno(11)
$$
and their relative phase is not zero in general.
Thus we can see that the off-diagonal interaction of the charginos with $Z$
breaks $CP$ at the tree level.

\vskip 1.0 true cm
\noindent{\bigbf 3.  {\bigit T}--odd Asymmetry}\bigskip

Our interest is in what $CP$--violating phenomena appear
in the chargino pair
production process of $e^+e^- \rightarrow \w2^+ \w1^-$
from the interaction in (9).  If one sums spins, $s_1$ and $s_2$, of both
$\w1^-$ and $\w2^+$, its cross section has no $CP$--violating term.
In other words one has to observe at least
the spin state of either of the charginos
to detect $CP$--violation effects.
For purposes of illustration, let us sum
$s_2$, but leave $s_1$ unsummed.  The spin $s_1$ can be chosen as
it is perpendicular to the interaction plane.  In the C.M. system of
$e^+e^-$, the cross section becomes
$$\eqalign{
&\sum_{s_2}d\sigma(e^+e^- \rightarrow \w2^+ \w1^-)/d\cos\theta =
             {8 \pi\alpha_{\rm em}^2 \over \sin^2 2\theta_W}
             {p\over \sqrt S}{1\over (S-M_Z^2)^2 + \Gamma_Z^2M_Z^2}\cr
&\bigl[ (f_L^2 + f_R^2)\left(
              (|\GL12|^2 + |\GR12|^2)(E_1E_2 + p^2 \cos^2\theta)
              + 2\Re(\GL12{\GR12}^*) \mw1\mw2  \right) \cr
            & + (f_L^2 - f_R^2)\left(
              (|\GL12|^2 - |\GR12|^2) {\sqrt S}p\cos\theta
             +  2\Im(\GL12{\GR12}^*){\rm sign}(s_1)\mw2 p\sin\theta
       \right)\bigr],\cr
}\eqno(12)
$$
where $p$ is the magnitude of the chargino momentum,
$E_{1, 2}$ are the energies of ${\w1}_{ ,2}$,
$f_{L, R}$ are the coupling constants of the electron neutral current,
$f_L = 1/2 - \sw2$, $f_R = - \sw2$, and
sign$(s_1)$ = sign(${\bf s_1\cdot (p^-\times p_1))}$.

The last term in (12) breaks $T$--invariance,
since the final state of
sign$(s_1)>0$ and that of
sign$(s_1)<0$
are transformed to each other by time reversal.
The difference of the cross section between these two states
manifests $T$--violation at the first-order perturbation.
The asymmetry of these two states
$$
  A_T = {d\sigma({\bf s_1\cdot (p^-\times p_1)}>0) -
         d\sigma({\bf s_1\cdot (p^-\times p_1)}<0) \over
         d\sigma({\bf s_1\cdot (p^-\times p_1)}>0) +
         d\sigma({\bf s_1\cdot (p^-\times p_1)}<0)},
\eqno(13)
$$
would quantify how large the $T$--violation is.

For a numerical example, we show the result of a parameter set of
$\tb = 2$, $\m2 = 200$ GeV, and $\mH = 200 {\rm e}^{i\pi/4}$ GeV,
which leads to the chargino masses 133, 275 GeV
and the neutralino masses 83, 145, 203, 278 GeV.  In this parameter set,
if the squarks are as heavy as 3 TeV, the EDM of
the neutron can be as low as $0.9 \times 10^{-25} e\cdot$cm.
The figure 1(a) shows $A_T$ as a function of
the scattering angle $\theta$ at
$\sqrt S = 500$ GeV.  The figure 2(a)
shows $A_T$ as a function of $\theta$ and
$\sqrt S$.  The magnitude of $A_T$ is of order $10^{-2}$.
The magnitude is somewhat smaller than what one might expect
as a $T$--violation effect at the tree level.
This is due to the fact that
the electron neutral current is almost pure axial, i.e.
$|f_L| \simeq |f_R|$.
If one wishes a larger $A_T$, the polarized electron beam should be utilized.
The electron beam is assumed to be left-handedly
polarized in Fig. 1(b) and
Fig. 2(b).  The asymmetry becomes as large as ${\cal O}(10^{-1})$
in this case.  From Fig. 2 it is  seen that the asymmetry becomes
saturated as $\sqrt S$ becomes large.
As long as $\sqrt S$ is sufficiently
larger than the $\w1$-$\w2$ threshold energy,
$A_T$ is about the same in
increasing the electron beam energy further.

\vskip 1.0 true cm
\noindent{\bigbf 4.  Discussion}\medskip

We implicitly assumed in the preceding section
that the spin of $\w1$ can be measured,
which is possibly a difficult thing to do.
What we can observe are the decay products of $\w1$.
Since their momenta are affected by the spin state of $\w1$,
they could substitute for $s_1$.
Letting $p_D$ denote a momentum of one of
the decay products,  we can then consider a $T$--odd asymmetry
$$
  A'_T = {d\sigma({\bf p_D\cdot (p^-\times p_1)}>0) -
          d\sigma({\bf p_D\cdot (p^-\times p_1)}<0) \over
          d\sigma({\bf p_D\cdot (p^-\times p_1)}>0) +
          d\sigma({\bf p_D\cdot (p^-\times p_1)}<0)}.
\eqno(14)
$$
The magnitude of $A'_T$ would be the same order as $A_T$.
But its $\sqrt S$
dependence will certainly be different from $A_T$,
because, as $\sqrt S$ becomes larger, $E_1$ becomes larger, and ${\bf p_D}$
orients to the direction of ${\bf p_1}$.  Since the $T$--violation term is
proportional to ${\bf p_D\cdot (p^-\times p_1)}$, $A'_T$ would come to
decrease at larger $\sqrt S$, and have a peak at some value of $\sqrt S$.
It may even not be possible to measure the momentum of $\w1$.
In this case the momentum of a decay product of $\w2$ could
substitute for $p_1$.

Since the diagonal coupling constants
of the charginos to $Z$ are real,
$CP$--violation does not occur at the tree level
in the production of the same mass eigenstates of the charginos.
However, the electric and the weak "electric" dipole moments of the charginos,
$D_{\wi}$, are generated at
the one-loop level, which break $T$ invariance.
These dipole moment terms give rise to $T$--odd asymmetry
in the same chargino pair-production processes.  From the dimensional grounds,
$A_T$ can be roughly estimated to be
$A_T \sim {\sqrt S}D_{\wi}/e$.
The main contributions to
$D_{\wi}$ are given by the loop diagrams involving the top quarks,
the top squarks, the $W$ bosons, and/or the Higgs bosons,
and roughly $D_{\wi}\sim 10^{-20} e\cdot$cm.  Thus
we could expect
$A_T \sim 10^{-4}$
in $e^+e^- \rightarrow \wi \bar{\wi}$,
which would be too small for detection
in the next $e^+e^-$ experiments.

Finally we should comment on the selectron exchange diagrams
which have been entirely ignored in discussing the chargino pair production.
They also contain $CP$--violating couplings, and produce $T$-odd asymmetry
at the tree level in the production of the different chargions.
However, the mass of the selectron should be as large as 3 TeV
according to the analysis of the electron EDM
if the charginos are as light as can be pair-produced at $\sqrt S = 500$GeV.
Thus the contribution of the selectron could be neglected
in this context.

In this report we have viewed $T$--odd asymmetry
in the chargino pair production processes.
If the imaginary phases of the supersymmetric parameters would have
their natural value of ${\cal O}(1)$,
the $CP$ violation originating from the supersymmetric standard model
could lead to measurable $CP$-violating phenomena.
If the charginos are produced in $e^+e^-$ collisions,
$T$-odd asymmetry would be observed in the angular distribution
of the final decay products.
\vskip 2.0 true cm

{\bigbf References}
\medskip
\item{[1]} J. Ellis, S. Ferrara, and D.V. Nanopoulos, {\it Phys.
           Lett.} {\bf B114} (1982) 231;
\item{ }   S.P. Chia and S. Nandi, {\it Phys. Lett.} {\bf B117} (1982) 45;
\item{ }   W. Buchm\"uller and D. Wyler, {\it Phys. Lett.} {\bf B121}
           (1983) 321;
\item{ }   J. Polchinski and M.B. Wise, {\it Phys. Lett.} {\bf B125}
           (1983) 393;
\item{ }   F. del \'Aguila, M.B. Gavela, J.A. Grifols, and
           A. M\'endez, {\it Phys. Lett.} {\bf B126} (1983) 71.
\item{[2]} D.V. Nanopoulos and M. Srednicki, {\it Phys. Lett.} {\bf B128}
           (1983) 61;
\item{ }   F. del \'Aguila, J.A. Grifols, A. M\'endez, D.V.
           Nanopoulos, and M. Srednicki, {\it Phys. Lett.} {\bf  B129}
           (1983) 77;
\item{ }   J.-M. Fr\`ere and M.B. Gavela, {\it Phys. Lett.} {\bf  B132}
           (1983)107;
\item{ }   E. Franco and M. Mangano, {\it Phys. Lett.} {\bf B135} (1984) 445;
\item{ }   J.-M. G\'erard, W. Grimus, A. Raychaudhuri, and
           G. Zoupanos, {\it Phys. Lett.} {\bf B140} (1984) 349;
\item{ }   J.-M. G\'erard, W. Grimus, A. Masiero, D.V. Nanopoulos,
           and A. Raychaudhuri, {\it Nucl. Phys.} {\bf B253} (1985) 93.
\item{[3]} M. Dugan, B. Grinstein, and L. Hall, {\it Nucl. Phys.} {\bf B255}
           (1985) 413;
\item{ }   T. Kurimoto, {\it Prog. Theor. Phys.} {\bf 73} (1985) 209.
\item{ }   A.I. Sanda, {\it Phys. Rev.} {\bf D32} (1985) 2992.
\item{[4]} S.T. Petcov, {\it Phys. Lett.} {\bf B178} (1986) 57.
\item{ }   P. Nath, {\it Phys. Rev. Lett.} {\bf 66} (1991) 2565.
\item{[5]} I.S. Altarev et al., {\it JETP Lett.} {\bf 44} (1986)
           460;
\item{ }   K.F. Smith et al., {\it Phys. Lett.} {\bf B234} (1990)
           191.
\item{[6]} S.A. Murthy et al., {\it Phys. Rev. Lett.} {\bf 63}
           (1989) 965;
\item{ }   D. Cho et al., {\it Phys. Rev. Lett.} {\bf 63}
           (1989) 2559;
\item{ }   K. Abdullah et al., {\it Phys. Rev. Lett.} {\bf 65}
           (1990) 2347.
\item{[7]} Y. Kizukuri and N. Oshimo, {\it Phys. Rev.} {\bf D45} (1992) 1806;
           {\bf D46} (1992) 3025.
\item{[8]} Y. Kizukuri, {\it Phys. Lett.} {\bf B193} (1987) 339.
\item{[9]} N. Oshimo, {\it Z. Phys.} {\bf C41} (1988) 129;
           {\it Mod. Phys. Lett.} {\bf A4} (1989) 145.
\item{[10]}Y. Kizukuri and N. Oshimo, {\it Phys. Lett.} {\bf B249}
           (1990) 449.
\item{[11]}Y. Kizukuri and N. Oshimo,
           in {\it Proceedings of the Workshop '$e^+e^-$ Collisions
           at 500 GeV:  The Physics Potential'}, ed. P.M. Zerwas,
           DESY 92-123B (1992).
\item{[12]}E.C. Christova and M.E. Fabbrichesi, {\it Phys. Lett.} {\bf B218}
           (1989) 470;
\item{ }   N. Oshimo, {\it Phys. Lett.} {\bf B227} (1989) 124.
\item{[13]}For a recent review of the MSSM, see e.g. H.E. Haber,
           Santa Cruz Institute for Particle Physics preprint,
           SCIPP 92/33 (1993), and references therein.
\vfill\eject

\nopagenumbers
\vsize =27 true cm

\topinsert
\vskip 24.0 true cm
\includegraphics{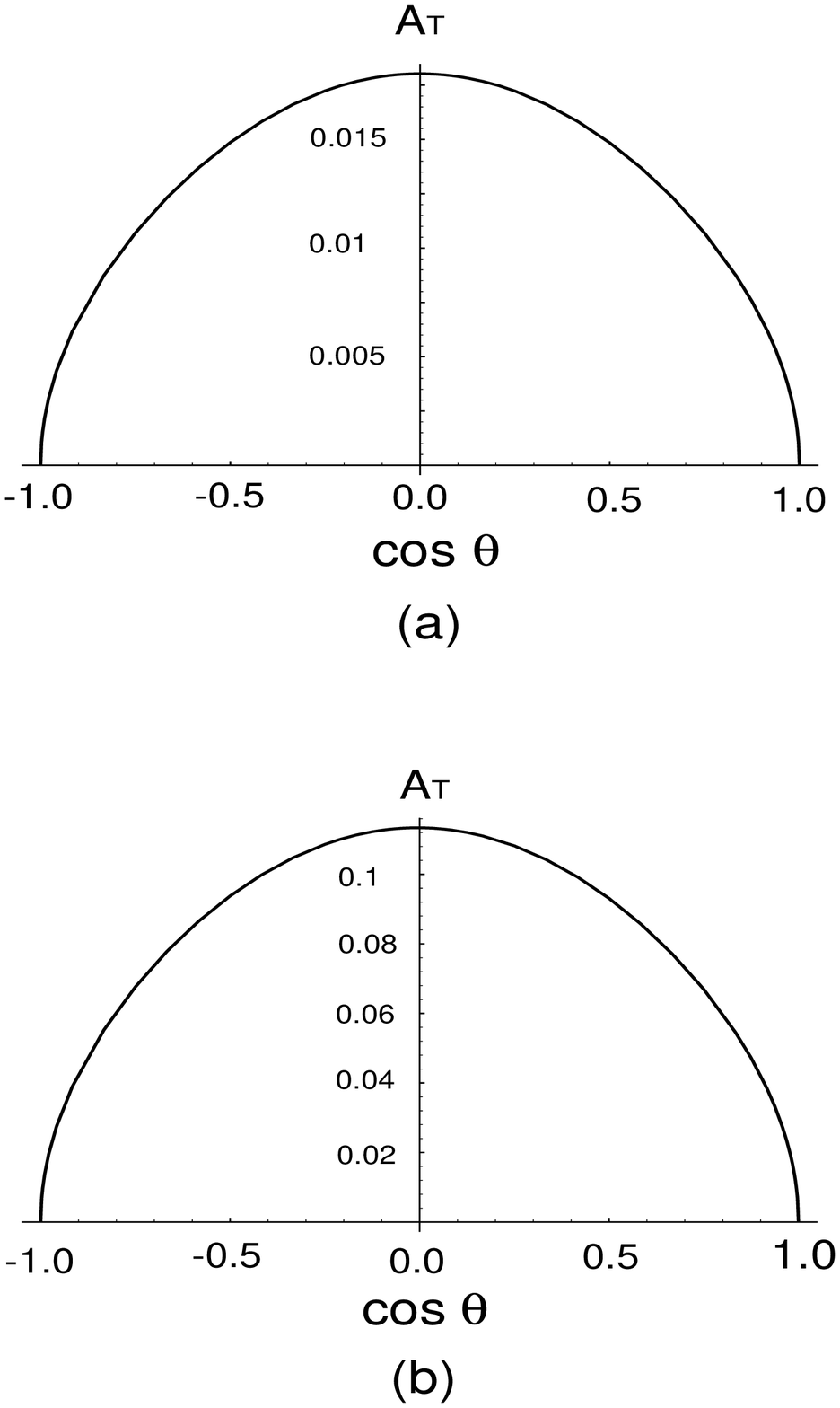}
\item{Fig. 1}$T$--odd asymmetry as a function of $\cos\theta$ at
             $\sqrt S = 500$ GeV;
             (a) unpolarized electron beam,
             (b) left-handed polarized electron beam.
\endinsert
\vfill\eject

\topinsert
\vskip 24.0 true cm
\includegraphics{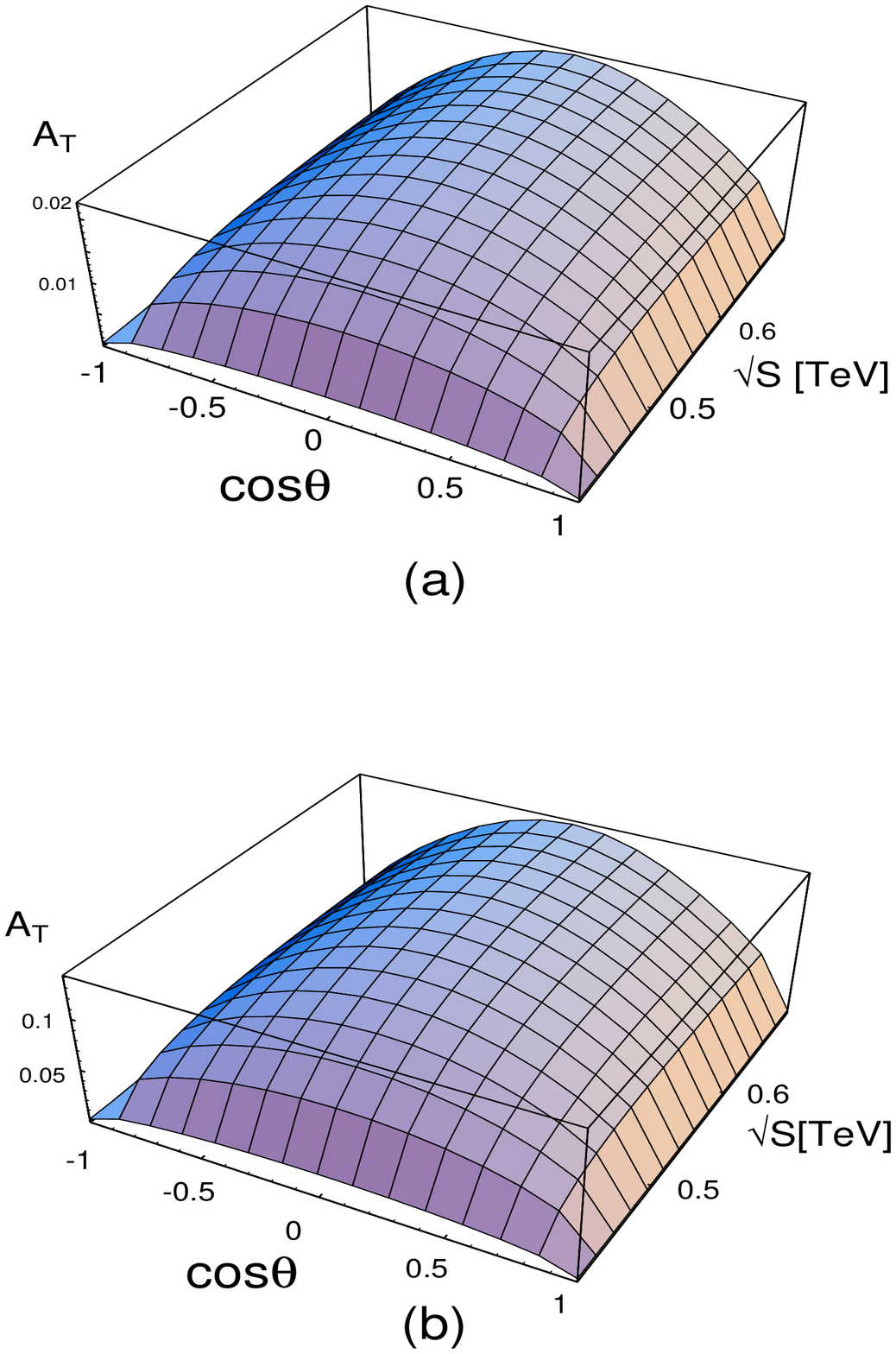}
\item{Fig. 2}$T$--odd asymmetry as a function of $\cos\theta$ and $\sqrt S$;
             (a) unpolarized electron beam,
             (b) left-handed polarized electron beam.
\endinsert
\vfill\eject
\end